\documentclass[aps,prd,groupaddress,showkeys]{revtex4}

\usepackage{epsfig}
\usepackage{amsmath}
\usepackage{amssymb}
\usepackage{color}
\usepackage{array}

\newcommand{\slabel}[1]{\label{sec:#1}}
\newcommand{\elabel}[1]{\label{eqn:#1}}
\newcommand{\flabel}[1]{\label{fig:#1}}

\newcommand{\cref}[1]{\ref{cpt:#1}}
\newcommand{\sref}[1]{\ref{sec:#1}}
\newcommand{\eref}[1]{\ref{eqn:#1}}
\newcommand{\fref}[1]{\ref{fig:#1}}

\newcommand{\listBegin}{\begin{tabular}{cp{4.5in}}}
\newcommand{\listEnd}{\end{tabular}}

\newcommand{\matrixBegin}[1]{\left[\!\!\left[ \begin{array}{#1}}
\newcommand{\matrixEnd}{\end{array} \right]\!\!\right]}

\newcommand{\fun}[2]{\,{#1}\!\left( {#2} \right)}

\newcommand{\imag}[1]{\fun{\mathcal{I}m}{#1}}
\newcommand{\real}[1]{\fun{\mathcal{R}e}{#1}}

\newcommand{\partderiv}[2]{\frac{\partial{#1}}{\partial{#2}}}
\newcommand{\partderivT}[2]{\partial{#1}/\partial{#2}}
\newcommand{\partderivS}[2]{\frac{\partial^2{#1}}{\partial{#2}^2}}

\newcommand{\dint}[3]{\int_{#1}^{#2} \hspace{-2ex} #3 ~}

\newcommand{\unit}[1]{\,\mbox{#1}}

\newcommand{\Hz}{\unit{Hz}}
\newcommand{\kHz}{\unit{kHz}}

\newcommand{\meter}{\unit{m}}
\newcommand{\cm}{\unit{cm}}

\newcommand{\um}{\unit{$\mu$m}}

\newcommand{\kg}{\unit{kg}}

\newcommand{\GPa}{\unit{GPa}}

\newcommand{\GJ}{\unit{GJ}}
\newcommand{\Kelvin}{\unit{K}}

\newcommand{\Watt}{\unit{W}}

\newcommand{\ppm}{\unit{ppm}}

\def\mtant{{\mbox{Ta}_2\mbox{O}_5}}
\def\tant{$\mtant$}
\def\sil{$\mbox{SiO}_2$}

\def\kB{k_B}
\def\dT{\delta T}
\def\rG{r_G}
\def\rP{r_\perp}

\def\hwave{$1/2$-wave}
\def\qwave{$1/4$-wave}

\def\prat{\sigma}

\def\abar{\bar{\alpha}}

\def\bbar{\bar{\beta}}
\def\rbar{\bar{r}}

\def\bulk{{s}}
\def\coat{{c}}

\def\dzp{\partial_\phi^z}

\def\abb{\abar_\bulk}

\def\abc{\abar_\coat}

\def\dab{\Delta \abar}

\def\phib{\phi_\bulk}
\def\phic{\phi_\coat}
\def\CcCb{\frac{C_\coat}{C_\bulk}}
\def\CzCb{\frac{C(z)}{C_\bulk}}
\def\CrCb{\frac{C(\vec{r})}{C_\bulk}}
\def\IrP{I(\rP)\,}
\def\srcwt{T F_0 \sin( \omega t ) \,}

\def\rdel{r_T}
\def\rrdel{\sqrt{2} \rdel}
\def\rdc{r_{T\coat}}
\def\rdb{r_{T\bulk}}
\def\rdi{\gamma}
\def\rdic{\gamma_\coat}
\def\rdib{\gamma_\bulk}
\def\xic{\xi}
\def\xicos{\cos(\xi)}
\def\xicosh{\cosh(\xi)}
\def\xisin{\sin(\xi)}
\def\xisinh{\sinh(\xi)}

\def\rhoTE{\rho_{TE}}
\def\rhoTR{\rho_{TR}}
\def\brhoTR{\bar{\rho}_{TR}}
\def\pE{p_E}
\def\pR{p_R}

\def\StoZ{S_{TO}^{\Delta z}}
\def\StoT{S_{TO}^{\Delta T}}

\def\SZ{S^{\Delta z}}

\def\dintz{\dint{0}{\infty}{dz}}

\usepackage{graphicx}

\begin{document}


\title{Thermo-optic noise in coated mirrors for high-precision optical measurements}
\author{M.Evans,
 S.Ballmer,
 M.Fejer,
 P.Fritschel,
 G.Harry,
 G.Ogin}


\begin{abstract}
Thermal fluctuations in the coatings used to make high-reflectors
 are becoming significant noise sources in precision optical measurements
 and are particularly relevant to advanced gravitational wave detectors.
There are two recognized sources of coating thermal noise,
 mechanical loss and thermal dissipation.
Thermal dissipation causes thermal fluctuations in the coating which produce
 noise via the thermo-elastic and thermo-refractive mechanisms.
We treat these mechanisms coherently, give a correction for finite coating
 thickness, and evaluate the implications for Advanced LIGO.
\end{abstract}

\maketitle

\section{Introduction and Main Result}

Thermal fluctuations in the coatings used to make high-reflectors
 are becoming significant noise sources in
 precision optical measurements \cite{Numata04}\cite{Ludlow07}\cite{Marquardt08}.
Though masked by other noise sources in the currently operating
 first generation interferometric gravitational-wave antennae
 (e.g., GEO \cite{httpGEO}, LIGO \cite{httpLIGO},
  TAMA \cite{httpTAMA}, Virgo \cite{httpVirgo}),
 designers of second generation gravitational-wave antennae expect
 coating thermal noise to be the dominant noise source in the detector's
 most sensitive frequency band \cite{Fritschel2003}.
Reduction of coating thermal noises has the potential to significantly
 increase the sensitivity, and thus the detection rate, of these
 large scale detectors.

Coating thermal noises are defined by differences between the coating
 material and the substrate material\footnote{
 To be precise, coating Brownian noise is simply the
  coating's contribution to the total Brownian noise,
  which is worth discussing separately because the coating
  materials typically have much higher loss than the substrate.
 Coating thermo-optic noise, on the other hand,
  is a phenomenon which results from the differences between
  the coating and the substrate.}.
There are two recognized sources of coating thermal noise,
 mechanical loss and thermal dissipation.
The first of these leads to ``coating Brownian'' noise, which,
 while not the topic of this paper, serves as measure against which we
 will compare our results \cite{Harry07}.
The second, thermal dissipation in the coating, leads to temperature fluctuations,
 which can cause ``thermo-optic'' noise via thermal expansion of
 the coating, and thermal change in refractive index
 of the coating material \cite{Brag00}.

Despite their common origin, coating thermo-elastic and
 thermo-refractive noises have not been treated in a coherent way \cite{Brag03}\cite{Levin08}.
Since the two mechanisms can be of the same order of magnitude
 a coherent treatment has the potential to greatly change the
 predicted magnitude of thermo-optic noise.

The purpose of this paper is to unify the thermo-optic mechanisms.
The formulaic result of this unification is presented
 later in this section, and derived in section \sref{ThermoDeriv}.
A correction for coatings of non-negligible thickness is given
 in \sref{ThickCoatCorr}.
In section \sref{aLIGO}, we evaluate the thermo-optic
 and Brownian noises expected to be present in Advanced LIGO,
 given current understanding of coating material parameters
 and detector design.
Finally, in the appendices we give equations for evaluating the
 average material constants of a multi-layer coating, we describe
 the dependence of the reflection phase of a coating on its temperature,
 and we relate coating thermo-elastic noise to substrate thermo-elastic noise.

The power spectrum of thermal fluctuations responsible for thermo-optic noise,
 as observed by a sensing beam with a Gaussian profile, is given by\cite{Levin08}
\begin{equation}
\StoT = \frac{2 \sqrt{2}}{\pi}
 \frac{\kB T^2}{\rG^2\sqrt{\kappa C \omega}}
 \elabel{StoT}
\end{equation}
(see the table at the end of this section for a list of symbols
 representing material parameters, their definitions and units)\footnote{
 In all uses of $\StoT$ herein, the material parameters
 refer to those of the substrate.}.

These thermal fluctuations result in fluctuations in the phase of a field
 reflected by a mirror's coating, which for the sensing beam are equivalent
 to changes in the inferred position of that mirror via the simple relation\footnote{
We use plane-wave propagation with phase evolution given by
 $E(x, t) =  E(0, 0) \exp \left( i \omega_0 t - i (2 \pi d_{prop} / \lambda + \phi)  \right)$,
 where $\omega_0 = 2 \pi c / \lambda$, $c$ is the speed of light,
 $t$ is the time of the measurement,
 $d_{prop}$ is the distance propagated,
 and $\phi$ is an additional phase due to propagation delay.
A mirror displacement $\Delta z$ shortens the propagation distance of the reflected field,
 and thus changes $d_{prop}$ by $-2 \Delta z$,
 while a change in reflection phase contributes directly to $\phi$.
Note that the mirror position $\Delta z$ is defined such than an expansion of the
 mirror coating results in a positive mirror position.}
\begin{equation}
\partderiv{~\Delta z}{T} = \partderiv{~\Delta z}{\phi} \partderiv{\phi}{T}
 = \frac{-\lambda}{4 \pi} \partderiv{\phi}{T}
 = \dzp \partderiv{\phi}{T},
 \elabel{dzdT}
\end{equation}
 where we define $\dzp \equiv -\lambda / 4 \pi$ to avoid repetition.

The spectral density of thermo-optic noise in a mirror's measured position
 is given by
\begin{equation}
\StoZ = \StoT ~ \left( \dzp \partderiv{\phic}{T} - \abb d \CcCb \right)^2
 \elabel{StoZ_phic}
\end{equation}
 which accounts for both thermo-optic mechanisms in $\partderivT{\phic}{T}$,
 the coating's overall reflection phase sensitivity to temperature\footnote{
 As a notational convention, $S^X_Y$ means ``the power spectrum of
 fluctuations in $X$ due to noise mechanism $Y$''.
 The units of this spectrum are the units of $X^2 / \Hz$.}.
Thermo-optic noise is explicitly limited to the coating by taking the difference
 between the temperature sensitivity of the coating and that of the substrate
 ($\partderivT{\phib}{T} = \dzp \abb$, integrated over the coating thickness $d$,
 and weighted by the relative heat capacity to give the term in equation \eref{StoZ_phic}).
Here, we will continue without diversion to
 an expression which can be easily evaluated,
 but to get a more precise result the thickness of the coating must
 be corrected for as described in section \sref{ThickCoatCorr},
 and the value of $\partderivT{\phic}{T}$ computed as described in
 appendix \sref{ReflCoat}.

Thermo-elastic expansion of the coating is complicated by the
 mechanical constraint of its attachment to the substrate \cite{Fejer03}.
Under the assumption that the coating elastic coefficients are similar
 to those of the substrate material, the effective thermal
 expansion coefficient simplifies to
\begin{equation}
\abc \sim 2 \alpha_\coat (1 + \prat_\coat).
\elabel{abar_simple}
\end{equation}
The same expression is valid for the constrained thermal expansion of the
 bulk material in a semi-infinite substrate, $\abb$.
A more complicated expression for $\abc$ applicable in the case of
 differing elastic coefficients is given in appendix \sref{CoatAvg}.

Both $\partderivT{n}{T}$ and thermal expansion play a role in changing
 the reflection phase of a coating, so while this mechanism is refereed
 to as ``thermo-refractive'', this is something of a misnomer.
Nonetheless, for a high-reflection coating made of \qwave\ doublets with a \hwave\ cap layer
 (see figure \fref{coating}),
 the thermo-refractive mechanism can be expressed in terms of the equivalent $\partderivT{n}{T}$
 of a hypothetical single layer of $n = 1$, $\alpha = 0$ material backed by a perfect reflector.
In this case, with a hypothetical layer of thickness $\lambda$, we find
\begin{equation}
\bbar \simeq \frac{B_H + B_L (2 (n_H / n_L)^2 - 1)}{4 (n_H^2 - n_L^2)},
\elabel{bbar}
\end{equation}
 where $B_X$ is the fractional change in optical path length with
 respect to temperature in material $X$
\begin{equation}
B_X = \beta_X + \abar_X n_X
\end{equation}
 with $X \in \{ L, H \}$ either the low-index material $L$,
 or the high-index material $H$.

\begin{figure}[h]
  \includegraphics[width=3.5in]{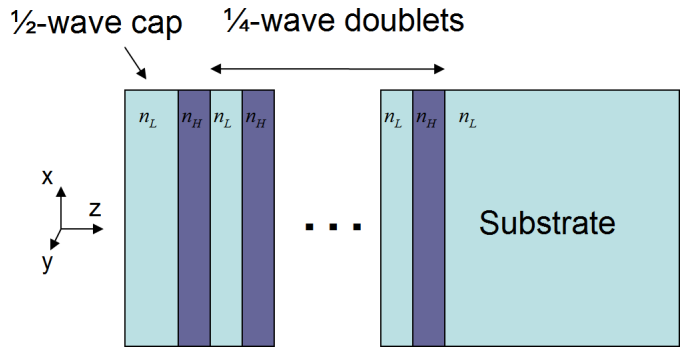}
  \caption{A high reflection coating made of \qwave\ doublets with a \hwave\ cap layer.}
  \flabel{coating}
\end{figure}

Accepting the approximations above, we can combine thermo-elastic and
 thermo-refractive (TE and TR) mechanisms to write the total coating
 reflection phase sensitivity to temperature as
\begin{equation}
\dzp \partderiv{\phic}{T} \simeq \abc d - \bbar \lambda,
\elabel{dzTO_dphi}
\end{equation}
 which allows us to rewrite (\eref{StoZ_phic}) as
\begin{equation}
\StoZ \simeq \StoT ~ \left( \abc d - \bbar \lambda - \abb d \CcCb \right)^2.
\elabel{StoZ_approx}
\end{equation}
Though this paper includes some refinements to previous works,
 the primary result is that the relative sign between the TE and TR
 mechanisms is \emph{negative}.

\subsection{Symbol Definitions}

The physical constants, material parameters,
 and frequently used symbols in this paper are:
\vspace{0.1in} \\
\begin{tabular}{clc}
 symbol & name  & SI unit \\
 \hline
 $\kB$ & Boltzmann's constant & $J / K$\\
 $T$ & mean temperature & $K$\\
 $\omega$ & angular frequency & $rad / s$\\
 $C$ & heat capacity per volume & $J / K m^3$\\
 $\kappa$ & thermal conductivity & $W / m K$\\
 $n$ & refractive index & $$\\
 $\alpha$ & thermal expansion & $1 / K$\\
 $\beta$ & $\partderivT{n}{T}$ & $$\\
 $E$ & Young's Modulus & $N / m^2$\\
 $\prat$ & Poisson ratio & $$\\
 $\lambda$ & beam wavelength & $m$\\
 $r_G$ & beam radius ($1/e^2$ power) & $m$\\
 $d$ & coating thickness & $m$\\
 $\dzp$ & $\partderivT{~\Delta z}{\phi} = -\lambda / 4 \pi$ & $m$\\
\end{tabular}
\vspace{0.1in} \\

Material parameters that appear with a subscript refer to either the
 bulk substrate material parameter, subscript $\bulk$,
 the average coating parameter, subscript $\coat$,
 or to one of the coating materials $L$, for low-refractive index,
 or $H$, for high-refractive index.

Material parameters which appears without a subscript,
 but as a function of $\vec{r}$ take on the value of the material
 at the location $\vec{r}$.
Thus, $\abar(\vec{r})$ is $\abc$ when $\vec{r}$ describes a point in the coating,
 and $\abb$ for points in the in the substrate.

Bars are used above symbols to express an ``effective'' coefficient.
These coefficients have the same units as their bar-less counterparts
 and the same general meaning, though taken in a specific context.
For example, $\abar$ has the same units as $\alpha$, and is a thermal
 expansion coefficient, but only in the context of a semi-infinite medium.

\section{Reflection Phase Noise}
\slabel{ThermoDeriv}

This section will derive equation \eref{StoZ_phic} from the Fluctuation-Dissipation
 Theorem (FDT)\cite{Callen51}.
We'll start by performing the derivation of Fejer's result for thermo-elastic noise
 using Levin's simpler approach\cite{Fejer03}\cite{Levin08}.
The solution to the more general problem of thermo-optic noise
 is derived second, following the same path.

The thermal fluctuations which are the source of thermo-optic noise
 are important to optical measurements because they change the result
 of position measurements based on reflecting a field from a mirror.
The fields used in these measurements are well described by
 a normalized Gaussian intensity profile\footnote{The definition of the Gaussian
 beam radius used by \cite{Brag03}\cite{Levin08}\cite{Liu00},
 is a factor of $\sqrt{2}$ different from the definition
 used in this paper and elsewhere \cite{Fejer03}\cite{Siegman}.
This results in an additional factor of 2 in the numerator of the noise spectra
 equations with respect to \cite{Brag03}\cite{Levin08}\cite{Liu00}.}
\begin{equation}
\IrP = \frac{2}{\pi \rG^2} e^{-2 \rP^2 / \rG^2},
\end{equation}
 where $\rG$ is the beam radius, and $\rP^2 = x^2 + y^2$ is the radius
 perpendicular to the beam's propagation direction (along the $z$ axis).

To go from thermal fluctuations to measured displacement noise we
 return to the foundation of this analysis.
Our application of the FDT starts with a gedanken experiment in which
 we consider an oscillating power injection in a small volume $\delta V$
 located at $\vec{r}$
\begin{equation}
\frac{P}{\delta V} = \srcwt q(\vec{r}).
\elabel{dPdV}
\end{equation}
 where $F_0$ is an arbitrary scale factor,
 and $\omega$ is the frequency of interest\footnote{Previous authors
 start from pressure injection \cite{Fejer03}\cite{Levin98}\cite{Liu00}.
 Pressure is converted to strain in the material, and then to
 power injection via the thermo-elastic mechanism.
We essentially follow the path of Levin \cite{Levin08},
 who speaks of entropy injection, which is equivalent to energy
 injection as expressed in his equation 12.
The application of a derivative with respect to time converts
 energy injection to power injection, a well defined quantity
 even in the presence of diffusion.}.
The form factor $q(\vec{r})$ connects the measurement
 variable $\hat{z}$ to temperature
 fluctuations $\delta T(\vec{r}, t)$ in the mirror via
\begin{equation}
\hat{z} = \oint \delta T(\vec{r}, t) \, q(\vec{r})
\end{equation}
 where the integral is formally over all space,
 though the integrand is presumably zero outside the mirror and its coating.

Power injection leads to heat flow and thus dissipation
 as expressed by
 \begin{equation}
W =  \left\langle \oint ~ \frac{\kappa}{T} (\vec{\nabla} \dT)^2 \right\rangle,
\elabel{Wdiss}
\end{equation}
 where the average $\langle \dots \rangle$ is over cycles of the
 power injection.
Finally, the FDT relates this dissipation to the
 spectral density of noise in the associated measurement variable by
\begin{equation}
\SZ = \frac{8 \kB T W}{F_0^2}.
\elabel{SZ}
\end{equation}

In the next section, as a illustrative example, we will derive Fejer's result
 for thermo-elastic noise using the approach outlined above.
The same approach is applied to the more complicated problem of thermo-optic
 noise in section \sref{CoatingThermoOptic}.

\subsection{An Example: Coating Thermo-elastic Noise}
\slabel{ThermoElastic}

As a concrete example, we will first apply the above formalism to
 derive coating thermo-elastic noise in the absence of any
 thermo-refractive mechanism,
 (previously performed in \cite{Brag03}\cite{Fejer03}).
The thermo-elastic readout variable
\begin{equation}
\hat{z}_{TE} = \oint \delta T(\vec{r}, t) q_{TE}(\vec{r})
\elabel{zTE}
\end{equation}
 describes the sensing beam's averaging of the thermally induced
 displacement of points on the mirror's surface.
The solution to the non-trivial problem of thermal expansion of a thin
 coating on a semi-infinite substrate is presented in \cite{Fejer03},
 appears in our equation \eref{abar_avg}, and is contained in the
 effective thermal expansion coefficient $\abar$.
From this we can write simply
\begin{equation}
q_{TE}(\vec{r}) = I(\rP) \abar(\vec{r}),
\end{equation}
 which leads to a thermo-elastic power injection
\begin{equation}
\frac{P_{TE}}{\delta V} = \srcwt \IrP \abar(\vec{r}).
\elabel{dPdV_TE}
\end{equation}

To remove the component of power injection which results in little temperature
 gradient and thus little heat flow, we subtract the substrate contribution
\begin{eqnarray*}
\frac{1}{C(\vec{r})} \frac{P_{TE_\coat}}{\delta V} &=&
 \frac{1}{C(\vec{r})} \frac{P_{TE}}{\delta V}
 - \frac{1}{C_\bulk} \frac{P_{TE_\bulk}}{\delta V} \\
&=& \srcwt \IrP
 \left( \frac{\abar(\vec{r})}{C(\vec{r})} - \frac{\abar_\bulk}{C_\bulk} \right)
\end{eqnarray*}
 We can then recast this into the form of (\eref{dPdV}) as
\begin{equation}
\frac{P_{TE_\coat}}{\delta V} = \srcwt q_{TE_\coat}(\vec{r}),
\elabel{dPdV_TEc}
\end{equation}
 where we have identified the coating thermo-elastic readout form factor
\begin{equation}
q_{TE_\coat}(\vec{r}) = \IrP \left( \abar(\vec{r}) - \abb \CrCb \right)
\elabel{qTEc}
\end{equation}
 which is zero in the substrate by design\footnote{
 $q_{TE_\coat}(\vec{r})$ is zero in the substrate simply because,
 for any value of $\vec{r}$ in the substrate $\abar(\vec{r}) = \abb$
 and $C(\vec{r}) = C_\bulk$.
 The reason for this choice of form factor is explained in appendix
 \sref{SubstrateTE}.}.

To maintain the simplicity of this example, we will assume that the
 coating and substrate are uniform, and that the coating is of thickness $d$
 which is small with respect to the thermal diffusion length
\begin{equation}
\rdel = \sqrt{\frac{\kappa}{C \omega}}.
 \elabel{rdel}
\end{equation}
With this assumption, we can consider all energy to be generated
 in this thin layer at the surface of the substrate and then flow inward.
Integrating (\eref{dPdV_TEc}) over $z$ we compute the energy flux into the
 substrate to be
\begin{equation}
\frac{P_{TE_\coat}}{\delta A} = \srcwt \IrP \dab d
\end{equation}
where
\begin{equation}
\dab =  \abc - \abb \CcCb.
\elabel{delta_abar}
\end{equation}

In order to connect this heat injection to $W$ in (\eref{Wdiss})
 we solve the diffusion equation
\begin{equation}
C \partderiv{\delta T}{t} = \kappa \nabla^2 \delta T
\elabel{HeatDiff}
\end{equation}
 with the boundary condition that the injected energy flows inward
\begin{equation}
\frac{P_{TE_\coat}}{\delta A} =
 -\kappa_\bulk \left. \partderiv{\delta T}{z} \right|_{z=0},
\elabel{HeatBoundDelta}
\end{equation}
 which ignores the very small radiation loss, as in \cite{Levin08}.

Further assuming that $\rG \gg \rdel$, we can ignore diffusion
 in the transverse dimensions, which yields the solution
\begin{equation}
\partderiv{\delta T}{z} \simeq \frac{-T F_0}{\kappa_\bulk} e^{\frac{-z}{\rrdel}}
 \sin \left( \omega t - \frac{z}{\rrdel}\right) \IrP \dab d,
\end{equation}
 from which we can compute the power dissipation
\begin{eqnarray}
W_{TE_\coat} &\simeq&  \left\langle \oint ~ \frac{\kappa_\bulk}{T}
 \left( \partderiv{\delta T}{z} \right)^2 \right\rangle \nonumber \\
W_{TE_\coat} &\simeq&  \frac{T F_0^2}{2 \sqrt{2} \pi \rG^2 \kappa_\bulk} \rdel ( \dab d )^2.
\elabel{Wdiss_TE}
\end{eqnarray}
Finally, returning to (\eref{SZ}), we arrive at the coating thermo-elastic
 noise spectrum
\begin{eqnarray}
S_{TE_\coat}^{\Delta z} & = & \frac{2 \sqrt{2} \kB T^2}{\pi \rG^2 \sqrt{\kappa_\bulk C_\bulk \omega}}
 \left( \dab d \right)^2 \nonumber \\
 & = & \StoT \left( \dab d \right)^2
\elabel{SteZ}
\end{eqnarray}
 which is equal to that of \cite{Fejer03}, and \cite{Brag03} under their
 simplifying assumptions.

\subsection{Coating Thermo-optic Noise}
\slabel{CoatingThermoOptic}

The thermo-elastic noise described above assumes
 that the relevant readout variable is based on the position of the
 surface of the mirror.
Interferometric sensors are, however, actually sensitive to the
 reflection phase of a surface as well as its position.

In the case of reflection from a planar surface, the position and reflection
 phase are related simply by $\delta z = -\delta \phi ~ \lambda / 4 \pi = \dzp ~ \delta \phi$,
 but for multi-layer coatings the relationship can be more complicated.
To account for this we generalize (\eref{zTE})
 to yield the thermo-optic readout variable
\begin{equation}
\hat{z}_{TO} = \dzp \oint \delta T(\vec{r}, t) \, \IrP
 \left( \partderiv{\phi(z)}{T} -  \CzCb \partderiv{\phi_\bulk}{T} \right),
\elabel{zTO}
\end{equation}
 where as before we have subtracted the substrate contribution so as to
 remove the component of heat injection which can be handled adiabatically.
From this we identify the thermo-optic form factor
\begin{equation}
q_{TO}(\vec{r}) = \IrP \dzp
 \left( \partderiv{\phi(z)}{T} -  \CzCb \partderiv{\phi_\bulk}{T} \right),
\end{equation}
 which is, as before, zero in the substrate.

Plugging into (\eref{dPdV}), we get
\begin{equation*}
\frac{P_{TO}}{\delta V} = \srcwt \IrP
 \dzp \left( \partderiv{\phi(z)}{T}
 - \CzCb \partderiv{\phi_\bulk}{T} \right).
\end{equation*}
Again we assume that the coating and substrate are uniform,
 and $d \ll \rdel \ll \rG$, so we can integrate over $z$
 to get the energy flux
\begin{equation}
\frac{P_{TO}}{\delta A} = \srcwt \IrP
 \left( \dzp \partderiv{\phic}{T}
 - \abb d \CcCb \right)
\end{equation}
 where
\begin{equation}
\partderiv{\phic}{T} = \dint{0}{d}{dz} \partderiv{\phi(z)}{T}
\end{equation}
 is the overall reflection phase sensitivity of the coating
 to temperature, as described in appendix \sref{ReflCoat}.

Following the path used for (\eref{SteZ}) above, we arrive at
\begin{equation}
W_{TO} \simeq  \frac{T F_0^2}{2 \sqrt{2} \pi \rG^2 \kappa_\bulk} \rdel
 \left( \dzp \partderiv{\phic}{T} - \abb d \CcCb \right)^2
\elabel{Wdiss_TO}
\end{equation}
 and thus
\begin{equation}
\StoZ = \StoT
 \left( \dzp \partderiv{\phic}{T} - \abb d \CcCb \right)^2
\elabel{StoZ_L}
\end{equation}
 which matches (\eref{StoZ_phic}).

\section{Thick Coating Correction}
\slabel{ThickCoatCorr}

Here we will allow for finite thickness coatings by
 removing the assumption that $d \ll \rdel$,
 while continuing to assume $\rG \gg \rdel$\footnote{To give
  some round numbers for gravitational-wave interferometers,
  $\rG \sim 5 \cm$ and $d < 10 \um$.
  For a \sil\ substrate, $\rdel \sim 40 \um$ around $100 \Hz$.}.
To do this we will need to solve the heat diffusion equation
 accounting for power deposition and diffusion in the coating.
Generalizing (\eref{HeatDiff}) to include a source term,
 but limiting heat flow to the $z$ axis
\begin{equation}
C \partderiv{\delta T}{t} = \kappa \, \partderivS{\delta T}{z}
 + \frac{P}{\delta L}
\elabel{HeatDiffSource}
\end{equation}
 with the one-dimensional power injection
 \begin{eqnarray}
\frac{P_{TO}}{\delta L} &=& \frac{1}{\IrP} \frac{P_{TO}}{\delta V} \\
 &=& \srcwt \left( \dzp \partderiv{\phi(z)}{T} - \abb \CcCb \right). \nonumber
\end{eqnarray}

We will approximate the thermo-optic power deposition in the coating
 with a constant thermo-elastic component, and a Dirac delta function for the
 thermo-refractive component
 since its effect is limited to the first few layers of the coating.
We can express this as
\begin{equation}
\frac{P_{TO}}{\delta L} \simeq \srcwt \left( \abc - \delta(z) \, \beta_{TR} - \abb \CcCb \right)
\end{equation}
 where we define
\begin{equation}
\beta_{TR} = \abc d - \dzp \partderiv{\phic}{T} \simeq \bbar \lambda
\end{equation}
 with $\partderivT{\phic}{T}$ and $\bbar$ as given in appendix \sref{ReflCoat}.

Following the method used in \cite{Fejer03},
 we transform (\eref{HeatDiffSource}) to a second-order
 differential equation in $z$
\begin{equation}
\theta(z) - \frac{1}{\rdi^2} \, \partderivS{\theta(z)}{z} = -\rho(z)
\elabel{HeatDiffImag}
\end{equation}
 where the relationships between the new and old variables are
\begin{eqnarray}
\elabel{Theta}
 \delta T(z, t) &=& \real{e^{i \omega t} \theta(z)} \\
 \frac{P(z, t)}{\delta L} &=& \omega C \, \real{ -i e^{i \omega t} \rho(z) } \nonumber \\
 \rdi &=& \sqrt{i \frac{\omega C}{\kappa}} = \frac{\sqrt{i}}{\rdel}. \nonumber
\end{eqnarray}
The homogeneous solutions to this equation in the coating and substrate are
\begin{eqnarray*}
 \theta_{h \coat}(z) &=& \theta_{d \coat} \cosh \left( \rdic z \right) \\
 \theta_{h \bulk}(z) &=& \theta_{d \bulk} \exp \left( -\rdib z \right)
\end{eqnarray*}
 where the coefficients $\theta_{d}$ will be determined
 by boundary conditions at $z = d$.
These equations satisfy the boundary conditions of no heat flow at
 $z = 0$ or $z = \infty$.

The particular solutions needed are for the two kinds of sources, TE and TR,
 both of which are limited to the coating.
The source terms are
\begin{eqnarray*}
 \rhoTE &=& \frac{T F_0}{\omega C_\coat} \left( \abc - \abb \CcCb \right)
  = \frac{T F_0}{\omega C_\coat} \dab \\
 \rhoTR(z) &=& -\delta(z) \, \frac{T F_0}{\omega C_\coat} ~ \beta_{TR}
  = \delta(z) \, \brhoTR
\end{eqnarray*}
 and the corresponding particular solutions are
\begin{eqnarray*}
 \theta_{pTE} &=& - \rhoTE \\
 \theta_{pTR}(z) &=& - \rdic ~ \brhoTR \exp(-\rdic z).
\end{eqnarray*}
We put all this together with boundary conditions at $z = d$
 that ensure continuity of temperature and conservation of energy
\begin{eqnarray*}
\theta_{h \bulk}(d) &=& \theta_{pTE} + \theta_{pTR}(d) + \theta_{hc}(d) \\
\kappa_\bulk \partderiv{}{z}\theta_{hs}(d) &=&
  \kappa_\coat \partderiv{}{z} \left( \theta_{pTR}(d) + \theta_{h \coat}(d) \right)
\end{eqnarray*}
 to find
\begin{eqnarray}
\theta_{d \coat} &=& \rhoTE + \brhoTR \rdic \exp(-\rdic d) (1 - R) / \psi_d \nonumber \\
\theta_{d \bulk} &=& -R \exp(\rdib d)
 \left( \rhoTE \sinh(\rdic d) + \brhoTR \rdic \right) / \psi_d \nonumber \\
\psi_d &=& \cosh(\rdic d) + R \sinh(\rdic d) \nonumber \\
R &=& \sqrt{\frac{\kappa_\coat C_\coat}{\kappa_\bulk C_\bulk}}
 = \frac{\kappa_\coat \rdic}{\kappa_\bulk \rdib}
 = \frac{\kappa_\coat \rdb}{\kappa_\bulk \rdc}.
\elabel{bound_R}
\end{eqnarray}

Before we lose ourselves among the equations,
 recall that our goal is to find the time averaged dissipation $W$,
 which is related to the temperature gradient in equation \eref{Wdiss}.
We now have $\theta(z)$ in hand, and (\eref{Theta}) relates this
 to $\delta T$, so our destination is near.
Summing the homogeneous and particular solutions to get $\theta(z)$,
 and taking the derivative with respect to $z$, we find
\begin{eqnarray*}
\partderiv{\theta_{\coat}(z)}{z} &=& \rdic (\theta_{d \coat} \sinh(\rdic z)
 + \rdic ~ \brhoTR \exp(-\rdic z)) \\
\partderiv{\theta_{\bulk}(z)}{z} &=& -\rdib \theta_{d \bulk} \exp \left( -\rdib z \right)
\end{eqnarray*}
From equations \eref{Wdiss} and \eref{Theta} we can see that
\begin{eqnarray}
W_{TO}^{thick} &\simeq& \frac{1}{\pi \rG^2} \left\langle \dintz \frac{\kappa}{T}
 \left( \partderiv{\delta T}{z} \right)^2 \right\rangle \nonumber \\
 &\simeq& \frac{1}{2 \pi \rG^2} \dintz \frac{\kappa}{T} \left| \partderiv{\theta(z)}{z} \right|^2
\end{eqnarray}
 where the transverse integrals over $\IrP^2$ have already been performed.

To arrive at a correction factor for thick coatings, we normalize the corrected
 thermo-optic dissipation above by that of a thin coating given in equation \eref{Wdiss_TO},
\begin{equation}
\Gamma_{tc} = \frac{W_{TO}^{thick}}{W_{TO}}
 = \frac{S_{TO}^{\Delta z_{thick}}}{\StoZ}.
\end{equation}
Taking the integral over the coating and substrate,
 we end with a complicated expression for the correction factor
\begin{eqnarray}
\Gamma_{tc} &=& \frac{\pE^2 \Gamma_0 + \pE \pR \xic \Gamma_1 + \pR^2 \xic^2 \Gamma_2}
 {R \xic^2 \Gamma_D } \\
\Gamma_0 &=& 2 (\xisinh - \xisin) + 2 R (\xicosh - \xicos) \nonumber \\
\Gamma_1 &=& 8 \sin(\xic/2) (R \cosh(\xic/2) + \sinh(\xic/2)) \nonumber \\
\Gamma_2 &=& (1 + R^2) \xisinh + (1 - R^2) \xisin + 2 R \xicosh \nonumber \\
\Gamma_D &=& (1 + R^2) \xicosh + (1 - R^2) \xicos + 2 R \xisinh \nonumber
\end{eqnarray}
 where we have made the following substitutions
\begin{eqnarray}
\pR = \frac{\brhoTR}{d \rhoTE + \brhoTR} &,&
\pE = \frac{d \rhoTE}{d \rhoTE + \brhoTR}
\end{eqnarray}
 using the dimensionless, frequency dependent, scale-factor
\begin{equation}
\xic = \frac{\sqrt{2} d}{\rdc} = \sqrt{\frac{2 \omega C_\coat}{\kappa_\coat}} d.
\end{equation}
Note that the power deposition fractions $\pE$ and $\pR$ can also be written as
\begin{eqnarray}
\pR = \frac{-\bbar \lambda}{\dab d - \bbar \lambda} &,&
\pE = \frac{\dab d}{\dab d - \bbar \lambda}.
\elabel{pRpE}
\end{eqnarray}
Applying this correction to equation \eref{StoZ_approx} gives
\begin{equation}
\StoZ = \StoT ~ \Gamma_{tc} \left( \dab d - \bbar \lambda \right)^2.
\elabel{StoZ_thick}
\end{equation}

For $d \ll \rdc$ or $\xic \ll 1$, we can use the much simpler expansion
\begin{eqnarray}
\Gamma_{tc} &\simeq& 1 + \frac{\pE^2 + 3 (\pR - R^2)}{3 R} \xic \nonumber \\
 && - \frac{\pE - 3 (1 - R^2)}{6} \xic^2
\end{eqnarray}
 which goes to $1$ as $\xic$ goes to $0$.
In the case of a very thick coating, with $d \gg \rdc$, the thermal fluctuations
 which generate noise via TE and TR mechanisms become independent,
 and thus they add in quadrature\footnote{Removing
 the thermo-refractive component (i.e., setting $\pE = 1, \pR = 0$),
 and normalizing by their $0^{th}$ order coefficient $R \xi / 2$,
 gives the results found in \cite{Fejer03}.}
\begin{eqnarray}
\Gamma_{tc} &\simeq& \frac{2 \pE^2}{R (1 + R) \xic^2} + \frac{\pR^2}{R}.
\end{eqnarray}
Thus, this correction expands our understanding beyond the simple notion
 that the TE and TR mechanisms have a relative negative sign.
Now we can say that
 TE and TR mechanisms have a relative negative sign if $d \ll \rdc$,
 are partially coherent and partially canceling if $d \sim \rdc$,
 and act as independent noises if $d \gg \rdc$.

\section{Implications for Advanced LIGO}
\slabel{aLIGO}

Having clarified the relationship between the thermo-optic mechanisms,
 a recomputation of the impact of this noise source is in order.
We will also take this opportunity to use the most recent
 information about the physical properties of the materials involved,
 and to apply an additional correction factor for the less-than-infinite
 size of the mirror.
To highlight the implications of this work,
 the results will be compared with Harry's result for coating Brownian noise \cite{Harry07}.

The Advanced LIGO mirrors are high reflectors with a multi-layer
 coating of alternating \sil\ and \tant.
The input mirrors will have a power transmission of $T = 1.4\%$
 with $\rG = 5.5 \cm$,
 while the end mirrors will have $T \simeq 5 \ppm$ with $\rG = 6.2 \cm$.
The mirrors are made of fused-silica, are $34 \cm$ in diameter and
 $20 \cm$ thick for a total mass of $40 \kg$.
In figures \fref{noiseITM} and \fref{noiseETM} we plot the coating
 related noises for coatings made of \qwave\ doublets.

\begin{figure}[h]
  \includegraphics[width=3.5in]{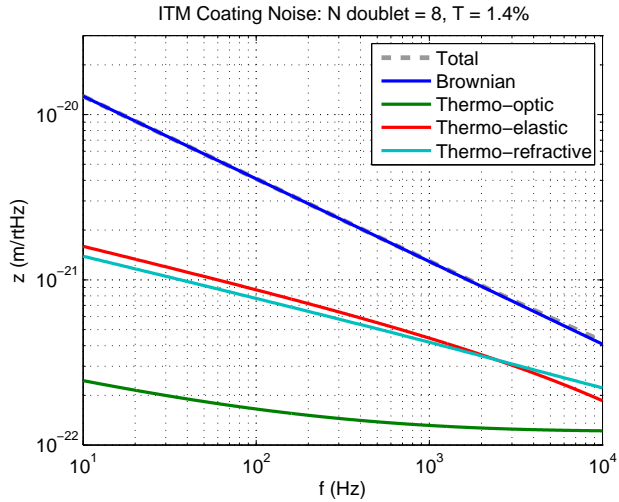}
  \caption{Thermo-optic noises and Brownian noise for an Advanced LIGO input mirror.}
  \flabel{noiseITM}
\end{figure}

\begin{figure}[h]
  \includegraphics[width=3.5in]{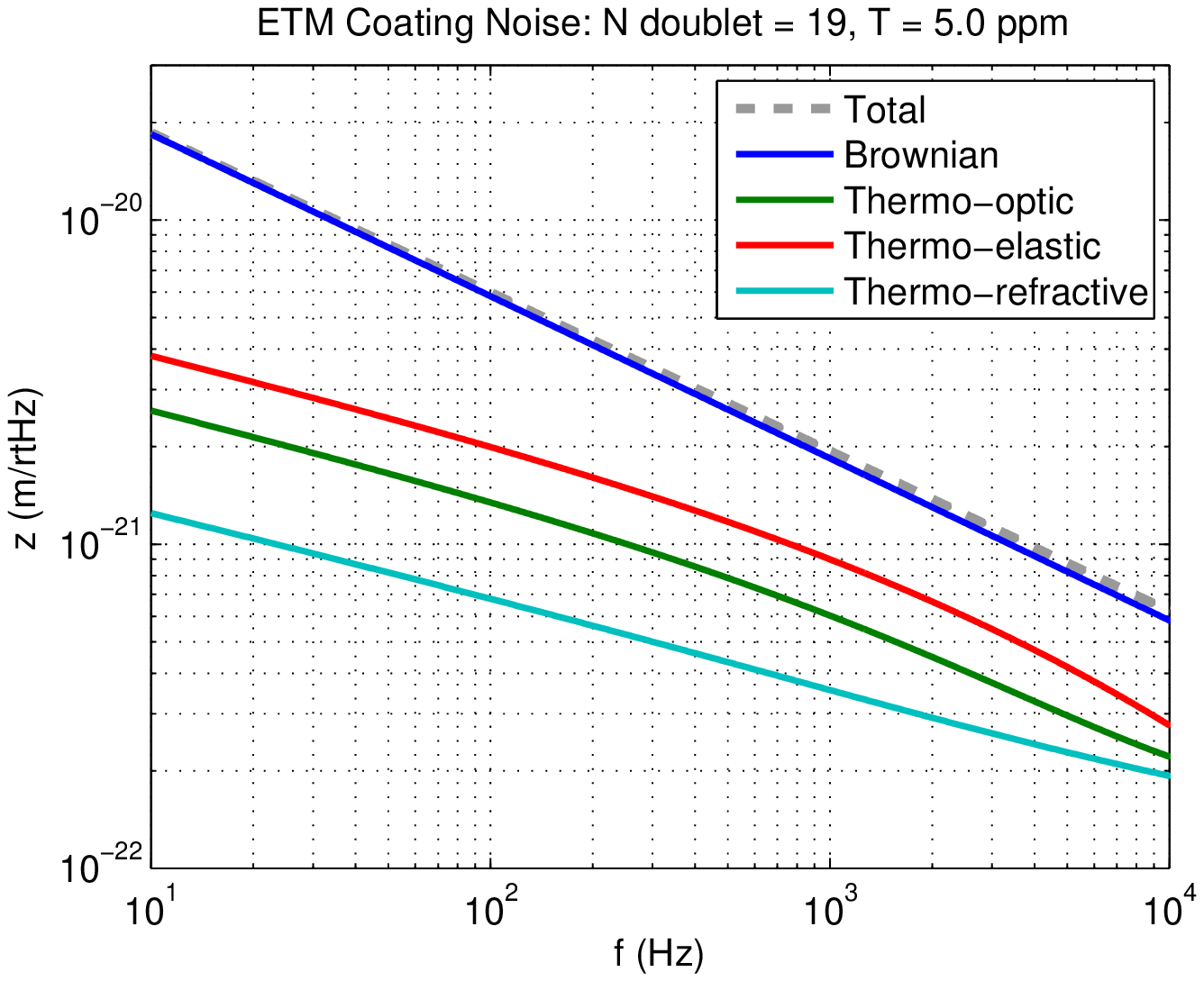}
  \caption{Thermo-optic noises and Brownian noise for an Advanced LIGO end mirror.}
  \flabel{noiseETM}
\end{figure}

\begin{figure}[h]
  \includegraphics[width=3.5in]{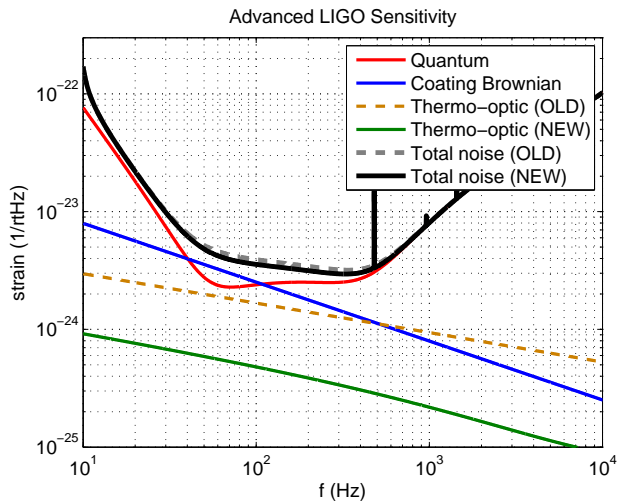}
  \caption{An Advanced LIGO sensitivity curve.  The thermo-optic curve labeled ``NEW''
   uses equation \eref{StoZ_full}, while the ``OLD'' curve uses a conservative estimate
   of TO noise: the sum of TR and TE, with the TE correction factor of 1.56 from \cite{Braginsky03-1}.}
  \flabel{aLIGOsens}
\end{figure}

We take the finite test-mass correction from \cite{Braginsky03-1} which,
 with the mirror and beam-size parameters given above, is $C_{fsm} \simeq 0.98$.
This multiplicative factor affects only the thermo-elastic mechanism,
\begin{equation}
\dab_{fsm} = C_{fsm} \dab,
\end{equation}
 as it represents a bending of the optic due to strains produced by the coating.
Adding this correction to equation \eref{StoZ_thick} gives
\begin{equation}
\StoZ = \StoT ~ \Gamma_{tc} \left( \dab_{fsm} d - \bbar \lambda \right)^2.
\elabel{StoZ_full}
\end{equation}
 where we use $\dab_{fsm}$ in (\eref{pRpE}) when computing $\Gamma_{tc}$.

Figure \fref{aLIGOsens} shows a representative Advanced LIGO sensitivity curve\footnote{
 To avoid clutter, many of the noise sources which form the ``Total'' curve are not shown in this
 figure \fref{aLIGOsens}.}.
While the difference between the result of equation \eref{StoZ_full} and a conservative
 estimate which simply takes the sum of the TR and TE mechanisms is less than 10\%,
 our coherent treatment of TO noise makes clear that it should not be considered a driving
 force in Advanced LIGO coating design.

It should also be noted that some of the material parameters used to
 make these figures are poorly constrained.
The thermal conductivity of \tant\ is simply assumed to match that of sapphire \cite{Fejer03}.
Fortunately this only effects the thick coating correction factor,
 and no reasonable value significantly changes the result below $1 \kHz$.
The value of $\beta$ for \tant\ is also poorly constrained,
 but again the range of tolerable values is large.
Thermo-optic noise remains below the conservative ``OLD'' curve in figure \fref{aLIGOsens}
 for values between $-10^{-4} / \Kelvin$ and $3 \times 10^{-4} / \Kelvin$.
The value of $\beta_\mtant$ used herein, from \cite{Gretarsson08},
 is comparable to previous values \cite{Cheng99}\cite{Tien00}.

The values of material parameters used for figures \fref{noiseITM},
 \fref{noiseETM} and \fref{aLIGOsens} are:
\vspace{0.1in} \\
\begin{tabular}{clc}
 symbol & \tant & unit \\
 \hline
 $\alpha$ & 3.6 & $10^{-6} / \Kelvin$\\
 $\beta$ & 14 & $10^{-6} / \Kelvin$\\
 $\kappa$ & 33 & $\Watt / \meter \Kelvin$\\
 $C$ & 2.1 & $\GJ / \Kelvin \meter^3$\\
 $E$ & 140 & $\GPa$\\
 $\prat$ & 0.23 & $$\\
 $n_H$ & 2.06 & $$\\
\end{tabular}
\vspace{0.1in} \\
\begin{tabular}{clc}
 symbol & \sil & unit \\
 \hline
 $\alpha$ & 0.51 & $10^{-6} / \Kelvin$\\
 $\beta$ & 8 & $10^{-6} / \Kelvin$\\
 $\kappa$ & 1.38 & $\Watt / \meter \Kelvin$\\
 $C$ & 1.64 & $\GJ / \Kelvin \meter^3$\\
 $E$ & 72 & $\GPa$\\
 $\prat$ & 0.17 & $$\\
 $n_L$ & 1.45 & $$\\
\end{tabular}
\vspace{0.1in} \\
These values are taken from \cite{Fejer03},
 with the exception of $\beta_\mtant$ noted above.

\section{Conclusion}

Thermo-optic noise results from thermal fluctuations in the coatings used
 to make high-reflection mirrors.
These thermal fluctuations affect the measured position of a mirror
 through the thermo-elastic and thermo-refractive mechanisms.
While both of these mechanisms have been known for some years,
 they were not treated coherently.
The coherent treatment presented herein shows that these two mechanisms
 appear with a relative \emph{negative} sign.
The effect is to essentially reduce thermo-optic noise to the point
 of insignificance for second generation gravitational-wave antennae.
While it is true that our current knowledge of the properties
 of coating materials is imprecise, it seems unlikely that better
 measurements, while desirable, will bring thermo-optic noise back
 into the realm of relevance.
This fact should help to guide coating research in the coming years.

\appendix

\section{Coating Average Properties}
\slabel{CoatAvg}

Optical coatings are made from alternating layers of materials with
 different refractive indices.
For properties other than the refractive index,
 as long as the length scales involved ($\rdel$ and $\rG$) are large
 compared to the layer thickness (typically $< \lambda / 2$),
 we can use suitably averaged material properties to
 represent the coating.
The equations given in this section are all taken from \cite{Fejer03},
 and are repeated here only for completeness and clarity.

The thermal expansion coefficient for a given layer $k$ in
 the coating is
\begin{equation}
\elabel{abar_k}
\abar_k = \alpha_k \frac{1 + \prat_\bulk}{1 - \prat_k}
 \left[ \frac{1 + \prat_k}{1 + \prat_\bulk} + (1 - 2 \prat_\bulk)
 \frac{E_k}{E_\bulk} \right]
\end{equation}
 and the volume average coefficient for a coating with $N$ layers
 each of thickness $d_k$ is
\begin{equation}
\elabel{abar_avg}
\abc = \sum_{k = 1}^{N} \abar_k \frac{d_k}{d}
\end{equation}
 where $d$ is the total coating thickness
\begin{equation}
\elabel{d_sum}
d = \sum_{k = 1}^{N} d_k.
\end{equation}

To compute the correction factor in section \sref{ThickCoatCorr}
 the average thermal properties of the coating are needed.
The heat capacity is a simple volume average,
\begin{equation}
\elabel{C_avg}
C_\coat = \sum_{k = 1}^{N} C_k \frac{d_k}{d}
\end{equation}
 while the average thermal conductivity involves the inverse
\begin{equation}
\elabel{K_avg}
\kappa_\coat = \left( \sum_{k = 1}^{N} \frac{1}{\kappa_k} \frac{d_k}{d} \right)^{-1} .
\end{equation}

\section{Reflection Phase of a Multi-Layer Coating}
\slabel{ReflCoat}

In this appendix we describe the method we use for computing the
 reflection phase of a multi-layer coating.
The initial discussion is somewhat pedantic, but it serves to give
 us a consistent notation which we develop in the subsections
 detailing the thermo-elastic and thermo-refractive mechanisms.

We start with the effective reflectivity of the interface between
 materials with refractive indices $n_1$ and $n_2$, passing from material
 $1$ to material $2$,
\begin{equation}
r_{1,2} = \frac{n_1 - n_2}{n_1 + n_2}.
\elabel{r_AB}
\end{equation}
Given two such transitions, from $1$ to $2$ and from $2$ to $3$,
 can equate the reflectivity to that of a two mirror cavity
\begin{equation}
r_{1,2,3} = \frac{-r_{2,1} + r_{2,3} ~ e^{-i \phi_2}}{1 - r_{2,1} r_{2,3} ~ e^{-i \phi_2}}
\elabel{r_cav}
\end{equation}
 where $\phi_2$ is the round-trip phase in material $2$.

Note that the reflectivity of the $2$ to $1$ transition appears
 in (\eref{r_cav}) with the indices in the order seen from inside the cavity.
In the following text we will use the relation $r_{k+1,k} = -r_{k,k+1}$,
 to keep the indices in increasing order, and then drop the second index,
 such that $r_k \equiv r_{k,k+1}$.

If we number the interfaces in our coating in the order of increasing depth
 (i.e., the coating layer in contact with the vacuum is $1$,
  and the layer in contact with the substrate is $N$)
 we can define a recursion relation using (\eref{r_cav})
\begin{equation}
\rbar_k = e^{-i \phi_k} \frac{r_k + \rbar_{k + 1}}
 {1 + r_k \rbar_{k + 1}}
\elabel{r_k}
\end{equation}
 where $\rbar_k = e^{-i \phi_k} r_{k,k+1,...,N}$ is the effective reflectivity
 of a coating layer, including the round-trip in that layer.
The base case for this recursion relation is the transition from the $N^{th}$ coating layer
 to the substrate,
\begin{equation}
\rbar_N = e^{-i \phi_N} r_{N,\bulk}
\end{equation}
 which can be evaluated with (\eref{r_AB}).

Extending our coating to include the external vacuum
 as layer $0$ provides a natural end to the recursion.
The reflectivity of the coating is then given by $r_\coat = \rbar_0$,
 and we can use $\phi_0$ to account for the overall expansion of the coating
 into the vacuum with $\phi_0 = \Delta_\coat / \dzp$,
 where $\Delta_\coat$ is the total change in coating thickness.

To use (\eref{r_k}) to compute changes in reflection phase
 one must take the derivative with respect to the round-trip phase in each layer.
Here we give the recursion relation and base case for these derivatives,
\begin{equation}
\partderiv{\rbar_k}{\phi_j} =
\begin{cases}
 e^{-i \phi_k} \frac{1 - r_k^2}{(1 + r_k \rbar_{k + 1})^2}
  \partderiv{\rbar_{k + 1}}{\phi_j} & k < j \\
 -i \rbar_k & k = j \\
 0 & k > j
\end{cases}
\elabel{dr_k}
\end{equation}
From the derivatives of the reflectivity of each layer,
 the derivative of the reflection phase of the coating as a whole is
\begin{equation}
\partderiv{\phic}{\phi_k}
 = \partderiv{\arg(\rbar_0)}{\phi_k}
 = \imag{\frac{1}{\rbar_0} \partderiv{\rbar_0}{\phi_k}}.
\elabel{phi_refl}
\end{equation}
For any quarter or half-wave coating,
 $\rbar_0$ is entirely real and its phase derivatives are entirely imaginary,
 so much of the apparent complexity is not real.

\subsection{Thermally Induced Changes}

For phase changes induced by a uniform change in temperature we have
\begin{equation}
\partderiv{\phic}{T} = \sum_{k=0}^N \partderiv{\phic}{\phi_k} \partderiv{\phi_k}{T}.
\elabel{phi_refl_T}
\end{equation}
The phase change due to thermo-elastic and thermo-refractive effects
 in a coating with layers of thickness $d_k$ are
\begin{eqnarray}
\partderiv{\phi_k}{T} &=& \frac{4 \pi}{\lambda} (\beta_k + \abar_k n_k) d_k
  = \frac{4 \pi}{\lambda} B_k d_k\nonumber \\
\partderiv{\phi_0}{T} &=& -\frac{4 \pi}{\lambda} \sum_{k=1}^N \abar_k d_k = \abc \frac{d}{\dzp}.
\elabel{phi_TO}
\end{eqnarray}
 where, as previously noted, we use $\phi_0$ to account for
 the overall expansion of the coating.
For any real coating, one can evaluate this expression numerically,
 and thus find $\partderivT{\phic}{T}$ for that coating.

\subsection{Relative Sign of TE and TR in \qwave\ Coatings}

Of particular interest are high-reflection coatings made
 of \qwave\ layers of alternating low-n and high-n material.
For simplicity, we'll assume that the high-n layers have $n_H > n_\bulk$
 and that the low-n layers have $n_L = n_\bulk$.
Thus, the reflectivity from high-n to low-n, is
\begin{equation}
r_H = \frac{n_H - n_L}{n_L + n_H}.
\elabel{r_H}
\end{equation}
As a transition from the vacuum, the first layer is of low-n material
 and \hwave\ in optical thickness, such that
\begin{equation}
r_0 = \frac{1 - n_L}{1 + n_L}.
\end{equation}
Summarizing, these coatings have the following properties
\begin{equation}
e^{-i \phi_k} =
\begin{cases}
~ 1 & k <= 1 \\
-1 & k > 1
\end{cases}
\nonumber
\end{equation}
\begin{equation}
r_k =
\begin{cases}
r_0 & k = 0 \\
r_H & k ~ \mbox{even} \\
-r_H & k ~ \mbox{odd} \\
\end{cases}
\nonumber
\end{equation}

From the above we can at least determine the signs of the
 various phase derivatives.
We start by noting that
\begin{equation}
\mbox{sign}(\rbar_k) =
\begin{cases}
-1 & k = 1 \\
-1 & k ~ \mbox{even} \\
~ 1 & k ~ \mbox{odd} \\
\end{cases}
\nonumber
\end{equation}
 and that (\eref{dr_k}) inverts the sign of the derivative
 for each layer with $k > 1$.
Even numbered layers start with $\partderivT{\rbar_k}{\phi_k}$ positive,
 experience $k - 2$ sign inversions, and thus end with a positive sign.
Odd numbered layers, on the other hand, start with $\partderivT{\rbar_k}{\phi_k}$ negative,
 experience an odd number of sign inversions, and thus these also end with a positive sign.
Since $\rbar_0$ is negative, we are ensured that
\begin{equation}
\mbox{sign} \left( \partderiv{\phic}{\phi_k} \right) = -1, ~ \mbox{for all $k$}.
\end{equation}
It follows that, for any high-reflection coating of this construction,
 thermo-elastic and thermo-refractive effects will appear with opposite
 sign in (\eref{phi_refl_T}), thanks to the relative minus sign in (\eref{phi_TO}).

\subsection{Approximation for High Reflectors}

While equations \eref{phi_refl_T} and \eref{phi_TO} are accurate and easy to use
 in numerical computation, they offer little intuitive understanding and fail
 to provide a concise expression for the thermo-optic mechanisms.
To address this, we give an approximation which is useful for high-reflection coatings.

The thermo-elastic mechanism, which arises from motion of the coating's
 surface, is accounted for by the $k = 0$ term in equation \eref{phi_refl_T}
 (also the second line in equation \eref{phi_TO}).
This term can be expressed in terms of the
 average coating expansion coefficient as
\begin{equation}
\partderiv{z_{TE}}{T} = \dzp \partderiv{\phi_0}{T} = \abc d
\end{equation}

The thermo-refractive mechanism is accounted for by the terms with $k > 0$
 in (\eref{phi_refl_T}), which can be thought of as the change in reflection phase
 as measured at a point on the coating's surface.
We define an effective TR coefficient $\bbar$ such that\footnote{
 We chose to make $\bbar$ positive and write explicitly the
 negative sign shown in the previous section.}
\begin{equation}
\partderiv{z_{TR}}{T} = -\bbar \lambda
\end{equation}
For a coating made entirely of \qwave\ doublets,
 $\bbar$ can be approximated by
\begin{equation}
\bbar_{QW} \simeq \frac{n_L^2 \bbar_H + n_H^2 \bbar_L}{4 (n_H^2 - n_L^2)},
\elabel{bbarQW}
\end{equation}
 as given in \cite{Brag00}.
A \qwave\ cap layer is, however, counter-productive and not used in high-reflectors.

To find $\bbar$ for the common HR coating
 (made of \qwave\ doublets with a \hwave\ cap layer),
 we modify $\bbar_{QW}$ by approximating $\rbar_k \simeq \mbox{sign}(\rbar_k)$.
Since the sign of $\rbar_0$ is minus in the \qwave\ case and plus in the \hwave\ case,
 each term in (\eref{dr_k}) with $j > 0$ is reduced by
\begin{equation}
\frac{(1 + r_0)^2}{(1 - r_0)^2} = \frac{1}{n_L^2}.
\end{equation}
Furthermore, we must include the additional \qwave\ of material in the thicker cap layer
\begin{equation}
\partderiv{\phi_{\coat,HW}}{\phi_1} = \frac{1 + r_0}{1 - r_0} = \frac{1}{n_L}
\end{equation}
 so that the additional temperature sensitivity is
\begin{equation}
 \bbar_L \partderiv{\phi_{\coat,HW}}{\phi_1} \frac{1}{4 n_L}
  = \bbar_L \frac{1}{4 n_L^2}.
\end{equation}

Putting these corrections together gives
\begin{equation}
\bbar \simeq \frac{\bbar_{QW}}{n_L^2} + \frac{\bbar_L}{4 n_L^2}
\end{equation}
 which can be rearranged to match equation \eref{bbar}.
Equation \eref{dzTO_dphi} arises simply from the sum of TE and TR terms
\begin{equation}
\dzp \partderiv{\phic}{T} = \partderiv{z_{TE}}{T} + \partderiv{z_{TR}}{T}
 \simeq \abc d - \bbar \lambda.
\end{equation}
For alternating layers of \sil\ and \tant, this approximation is within
 a few percent for coatings with more than $\sim 6$ doublets.

\section{Relationship to Substrate Thermo-Elastic Noise}
\slabel{SubstrateTE}

The spectrum of thermal fluctuations described by (\eref{StoT}),
 and derived previously in \cite{Brag03} and \cite{Levin08},
 can be rearranged with the help of the thermal diffusion length.
If we rewrite (\eref{StoT}) as
\begin{equation*}
\StoT = \frac{\sqrt{2} \kB T^2}{\omega C_\bulk \rdel^3} ~ \frac{2 \rdel^2}{\pi \rG^2 },
\end{equation*}
 we can see that the first fraction is the spectral density of the
 thermodynamic fluctuation in a volume defined by the diffusion length,
 while the second is the Gaussian beam average over these volumes.

The coating thermo-optic coupling is designed such that a similar equation applied
 to the substrate would result in zero.
The reason for this is that the loss associated with the coating results from
 non-adiabatic heat flow due to the difference between the coating and substrate.
The substrate thermo-elastic noise, on the other hand, results
 from adiabatic heat flow on the scale of the the beam radius $\rG$,
 and is thus smaller by a factor of $\sim \rdel / \rG$.
See, for instance, equation 2 in \cite{Brag03} which can be written in our notation as
\begin{eqnarray}
S_{TE_\bulk}^{\Delta z} &=& \frac{4 \kB T^2}{\sqrt{\pi} \omega C_\bulk \rG^3}
 ~ (\abb \rdel)^2 \nonumber \\
 &=& \StoT ~ \sqrt{2 \pi} \frac{\rdel}{\rG} (\abb \rdel)^2
\elabel{SteZs}
\end{eqnarray}

To give an idea of the relative importance of substrate and coating thermo-elastic noise,
 we divide the coating thermo-elastic noise in (\eref{SteZ}) by (\eref{SteZs}) and define
 the thermo-elastic ratio
\begin{eqnarray}
R_{TE} \equiv \frac{S_{TE_\coat}^{\Delta z}}{S_{TE_\bulk}^{\Delta z}} =
 \frac{d^2 \rG}{\sqrt{2 \pi} \rdel^3} \frac{\dab^2}{\abb^2}.
\end{eqnarray}
In the case of a gravitational-wave interferometers we have roughly,
 $\rG \sim 5 \cm$ and $d \sim 5 \um$.
For a fused silica substrate, $\rdel \sim 40 \um$ around $100 \Hz$,
 such that $R_{TE} \sim 10 (\dab / \abb)^2 \sim 150$,
 indicating that the substrate contribution is insignificant.
For a sapphire substrate ($\rdel \sim 130 \um$ and $\alpha_\bulk = 5.6 \times 10^{-6}$),
 on the other hand, the substrate contribution is dominant
 $R_{TE} \sim 0.2 (\dab / \abb)^2 \sim 0.1$.

\bibliography{ThermoOptic}

\end{document}